\documentclass[usenatbib]{mn2e}
\usepackage{graphicx}
\usepackage{times}
\usepackage{amssymb}
\usepackage{amsmath}
\usepackage{hyperref}

\graphicspath{{./figures/}}

\newcommand{\elem}[2]{$\mathrm{^{#2}#1}$}
\newcommand{\melem}[2]{\mathrm{^{#2}#1}}
\newcommand{\snia}{SN~Ia}
\newcommand{\snias}{SNe~Ia}

\def\apj{ApJ}
\def\apjs{ApJS}
\def\aap{A\&A}%
\def\mnras{MNRAS}%

\title[Monte Carlo radiation hydrodynamics]{Monte Carlo radiation hydrodynamics:
methods, tests and application to supernova Type Ia ejecta}
\author[Noebauer~et~al.]{U.~M.~Noebauer,$^1$\thanks{unoebauer@mpa-garching.mpg.de}
S.~A.~Sim,$^2$
M.~Kromer,$^{1}$
F.~K.~R\"opke$^{1,3}$
and W.~Hillebrandt$^{1}$\\
  $^1$Max-Planck-Institut f\"ur Astrophysik, Karl-Schwarzschild-Str.~1, 85748
  Garching, Germany\\
  $^{2}$  Research School of Astronomy \& Astrophysics, Mount Stromlo Observatory,
  Cotter Road, Weston ACT 2611, Australia\\
  $^{3}$ Institut f\"ur Theoretische Physik und Astrophysik, Universit\"at W\"urzburg,
  Emil-Fischer-Str.~31, D-97075 W\"urzburg, Germany
}

\begin{document}
\maketitle
\begin{abstract}
In astrophysical systems, radiation-matter interactions are important in
transferring energy and momentum between the radiation field and the surrounding
material. This coupling often makes it necessary to consider the role of
radiation when modelling the dynamics of astrophysical fluids.  During the last
few years, there have been rapid developments in the use of Monte Carlo methods
for numerical radiative transfer simulations. Here, we present an approach to
radiation hydrodynamics that is based on coupling Monte Carlo radiative transfer
techniques with finite-volume hydrodynamical methods in an operator-split manner.
In particular, we adopt an indivisible packet formalism to discretize the
radiation field into an ensemble of Monte Carlo packets and employ volume-based
estimators to reconstruct the radiation field characteristics. In this paper the
numerical tools of this method are presented and their accuracy is verified in a
series of test calculations.  Finally, as a practical example, we use our
approach to study the influence of the radiation-matter coupling on the
homologous expansion phase and the bolometric light curve of Type Ia supernova
explosions.
\end{abstract}

\maketitle

\section{Introduction}

In studying astrophysical objects, a detailed understanding and description of
the radiation field is vital, particularly if synthetic observables are to be
computed for comparison with observations.  Conceptually, the radiation field in
a fluid is not independent of the fluid state and their co-evolution has to be
described self-consistently within the framework of radiation hydrodynamics.
Depending on the dynamical importance of the radiation field and the strength of
the radiation-matter coupling, different strategies can be followed. If the
energy associated with the radiation field is negligible compared to the total
energy content, a de-coupled approach may be followed. For example, such a method
has been used for the determination of synthetic light curves and spectra for
Type Ia supernova (\snia) explosions around maximum light \citep[e.g.][]{Kasen2006,
Kromer2009, Jack2011}. In cases where the radiative terms are dynamically
important, however, a fully de-coupled treatment of the radiation field is not
possible.  For such applications a variety of different techniques have been
used to follow the co-evolution of the radiation-matter state.  In optically
thick environments, the radiation field is well-described by the diffusion
approximation and its evolution in radiation-hydrodynamical simulations can be
incorporated by using flux limited diffusion methods \citep{Levermore1981}. This
numerical prescription is used, for example, in the modelling of radiation
dominated accretion discs \citep[e.g.][]{Turner2003, Hirose2009}. In the
opposite case of low optical depth, the influence of the radiation field may be
treated by including a radiative cooling term \citep[e.g.][]{Townsend2009,
Marle2011}, as is often done in studies of stellar winds
\citep[e.g.][]{Garcia-Segura1996, Mellema2002}.  In the intermediate regimes
between the two extremes, a full radiation-hydrodynamical description of the
radiation-matter state is necessary, for example when accounting for convective
motions in studies of stellar atmospheres \citep[e.g.][]{Stein1998,
Asplund2000}, shock breakouts in supernovae \citep[e.g.][]{Blinnikov2000,
Hoeflich2009, Piro2010} or when studying interactions of stellar explosions with
circumstellar material \citep[e.g.][]{Fryer2010, kasen2010}.

In this paper we present the numerical methods and the application of a new
approach to radiation hydrodynamics that is based on Monte Carlo radiative
transfer techniques. A similar strategy has been pursued in the calculations
presented in \citet{Kasen2011}. Monte Carlo methods have already shown
tremendous success in pure radiative transfer applications
\citep[e.g.][]{Fleck1971, Abbott1985, Mazzali1993, Long2002, Carciofi2006,
Kasen2006, Harries2011}.  Within this probabilistic approach, complex
radiation-matter interaction physics can be simulated and problems with
arbitrary geometries can be addressed. Here we aim to extend the Monte Carlo
method to radiation-hydrodynamical calculations and explore its practicality for
modern astrophysical applications.

The focus of this paper is to present the theoretical and numerical foundations
and to verify the operation of our Monte Carlo radiation-hydrodynamical method.
We begin with a brief overview of the theoretical concepts that govern radiating
flows in Section \ref{sec:theory}, which is followed by an extensive description
of the numerical methods of our approach in Section \ref{sec:numerics}. The
physical accuracy and the computational feasibility of the techniques presented
here are assessed in Section \ref{sec:testing}, in which the results of a series
of test calculations are described. As a first application of the method in
astrophysical environments we report in Section \ref{sec:application} on our
investigation of \snias{} ejecta. In particular, we study the influence of the
radiation-matter coupling on the ejecta structure and the resulting effects on
the bolometric light curve during the homologous expansion phase. We summarize
our results and conclude in Section \ref{sec:discussion}.

\section{Theoretical Background}
\label{sec:theory}

To model environments in which a significant part of the total energy is
stored in the radiation field, one must deal with the
coupled evolution of the matter state and the radiation field. The former
changes due to external forces, gradients of the thermodynamic variables and
the radiation pressure acting on the matter. Such radiation pressure gradients
are the consequence of anisotropies in the radiation field whose temporal
evolution is driven by its interactions with the surrounding medium.  Generally,
these interactions are strongly dependent on the state of matter, i.e.\ on
its density, temperature and composition. The theory of radiating fluids
provides an adequate self-consistent description of the dynamical behaviour of
the radiation-hydrodynamical state of the coupled radiation field-matter system.
In the following section, we will give a brief outline of some important aspects
of this theory. In-depth discussions can be found in standard
textbooks, e.g.\ \citet{Mihalas1984}.

To describe the energy and the momentum of radiating fluids, the standard
hydrodynamical equations expressing conservation of momentum and energy are
extended by including additional source terms that account for the influence of the
radiation field. In astrophysical environments, the physical viscosity is
typically insignificant compared to the viscosity inherent to the numerical
schemes used to model fluid flows. Consequently, the ideal Euler
equations are commonly employed.
Modified by the influence of the radiation field, these take the form
\begin{flalign}
  &\rho \frac{\mathrm{D}}{\mathrm{D} t} \bmath{u} = \bmath{f} - \nabla P +
  \bmath{G}, &  
  \label{eq:rh_momentum_eq}\\
  &\rho \frac{\mathrm{D}}{\mathrm{D} t} e =
  \bmath{u} \cdot \bmath{f} - \nabla \cdot (P \bmath{u}) + c G^0. &  
  \label{eq:rh_energy_eq}
\end{flalign}
Mass conservation, expressed by the continuity equation 
\begin{eqnarray}
  \frac{\mathrm{D}}{\mathrm{D} t} \rho  + \rho (\nabla \cdot \bmath{u})= 0,
  \label{eq:continuum_eq}
\end{eqnarray}
is not affected by the radiation field. Here, $\rho$, $\bmath{u}$, $e$ and $P$
denote the fluid density, velocity, total energy and thermodynamic pressure,
respectively. Possible external forces are accounted for by the force density
$\bmath{f}$. The radiation field acts as an additional energy and momentum
source in the form of the radiation 4-force components $G^0$ and $\bmath{G}$
(see below). Note that we have formulated the equations in terms of substantial
derivatives
\begin{eqnarray}
  \frac{\mathrm{D}}{\mathrm{D} t} = \frac{\partial}{\partial t} + \bmath{u}
  \cdot \nabla,
  \label{eq:substantial_derivative}
\end{eqnarray}
which capture changes in the co-moving fluid frame. 

The $G$-terms essentially describe momentum and energy flows caused by an
imbalance of absorption and emission interactions between the radiation field
and its surrounding medium. Quantitatively, these interactions are characterised
by the opacity, $\chi(\bmath{x}, t; \bmath{n}, \nu)$, and the emissivity
$\eta(\bmath{x}, t; \bmath{n}, \nu)$ of the medium. These material functions
depend on the frequency~($\nu$) and the propagation direction~($\bmath{n}$) of
the radiation and will in general vary with time~($t$) and
position~($\bmath{x}$), since the radiation-matter interactions depend strongly
on the fluid state. The radiation force components can be specified in terms of
these material functions and the specific intensity, $I(\bmath{x}, t;
\bmath{n}, \nu)$:
\begin{flalign}
  &G^0 = \frac{1}{c}\int_{0}^{\infty} \mathrm{d} \nu \int
  \mathrm{d}\Omega~[\chi(\bmath{x}, t; \bmath{n}, \nu) I(\bmath{x}, t;
  \bmath{n}, \nu) - \eta(\bmath{x}, t; \bmath{n}, \nu], &
  \label{eq:rad_force_0}\\
  &G^i = \frac{1}{c}\int_{0}^{\infty} \mathrm{d} \nu \int \mathrm{d}\Omega~n^i
  \left[\chi(\bmath{x}, t; \bmath{n}, \nu) I(\bmath{x}, t; \bmath{n}, \nu) -
  \eta(\bmath{x}, t; \bmath{n}, \nu) \right]. & \label{eq:rad_force_i}
\end{flalign}
These can be understood as the net absorbed or emitted energy and momentum,
respectively. The temporal evolution of the radiation field itself is in turn 
driven by the interaction with the environment
\begin{flalign}
  &\left(\frac{1}{c}\frac{\partial}{\partial t} + \bmath{n} \cdot \nabla
  \right)
  I(\bmath{x}, t; \bmath{n}, \nu) =  & \nonumber\\ 
  &\qquad \eta(\bmath{x}, t; \bmath{n}, \nu) -
  \chi(\bmath{x}, t; \bmath{n}, \nu) I(\bmath{x}, t; \bmath{n}, \nu). &
  \label{eq:transfer_eq}
\end{flalign}
The combination of Equations~(\ref{eq:rh_momentum_eq}), (\ref{eq:rh_energy_eq}),
(\ref{eq:continuum_eq}), (\ref{eq:transfer_eq}) and an equation of state,
relating thermodynamic pressure with internal energy, provides the full set of
radiation-hydrodynamical equations that describe a radiating fluid.  In this
formulation, Equations~(\ref{eq:rad_force_0}) and (\ref{eq:rad_force_i})
describe how the energy and momentum transfer is obtained from the temporal
evolution of the radiation field. In the following, we present in detail
the numerical approach we developed to solve the radiation-hydrodynamical
problem formulated by this set of equations.

\section{Numerical Methods}
\label{sec:numerics}

To determine the state of a radiating fluid,
Equations~(\ref{eq:rh_momentum_eq}), (\ref{eq:rh_energy_eq}),
(\ref{eq:continuum_eq}), (\ref{eq:transfer_eq}) and the equation of state have
to be solved simultaneously.  Key to our approach is the application of a simple
operator splitting scheme \citep[see e.g.][for a detailed description of the operator
splitting technique]{LeVeque2002}. In this Godunov-splitting framework, the
temporal evolution is determined progressively, accounting for the pure
hydrodynamical effects and the influence of the radiation field independently
and in sequence.  Specifically, in each time step, a new radiation
hydrodynamical state is found by first performing a fluid dynamical calculation
neglecting all radiative influences.  For this step we employ a finite-volume
hydrodynamical scheme, namely the \textit{piecewise parabolic method}
\citep[\textit{PPM},][]{Colella1984}.  This is followed by a second step which
only accounts for the influence of the terms in the equations governing the
evolution of the radiation field and the fluid-dynamics terms are neglected. For
this part of the simulation, we carry out time-dependent Monte Carlo radiative
transfer calculations, which allow us to evaluate the radiation terms and use
them to update the hydrodynamical state.

The following sub-sections describe our scheme and the involved computational
methods. We begin with an outline of the hydrodynamics solver in
Section~\ref{sec:hydro}, followed by a detailed presentation of the Monte Carlo
radiative transfer techniques in Sections~\ref{sec:MC_technique} to
\ref{sec:noise}. The final step of updating the fluid state to account for the
influence of the radiation field is described in
Section~\ref{sec:rad_influence}.

\subsection{Hydrodynamical Calculation}
\label{sec:hydro}

The hydrodynamical sub-problem of the operator-split approach is solved with
the piecewise parabolic method of \citet{Colella1984}. This higher-order Godunov
scheme is based on reconstructing a continuous fluid state from a discrete
representation on a computational grid by a series of parabolas. In the spirit
of finite-volume approaches, Riemann problems are defined at the cell interfaces
by discontinuities in the fluid properties, which arise from integrating
the reconstructed fluid state over the domains of dependence, i.e.\ the regions
that can influence the interfaces. The solutions to the Riemann problems
determine the flux through the interfaces. By balancing the resulting in- and
outflows in the grid cells, the temporal evolution of the fluid state can be
calculated for one time step.  This second-order reconstruction scheme provides
higher accuracy over the traditional constant-reconstruction method of
\citet{Godunov1959}. The detailed implementation of PPM in our radiation
hydrodynamics code follows \citet{Edelmann2010}. After determining the new fluid
state, a Monte Carlo simulation is started to address the evolution of the
radiation field as the second half of the splitting scheme.

\subsection{Monte Carlo Techniques}
\label{sec:MC_technique}

In our Monte Carlo approach the radiation field is discretized 
into a large number of Monte Carlo quanta, hereafter referred to as {\it packets}.
Each of them carries a fraction of the radiation field energy and is propagated
through the medium. The propagation path is determined stochastically but in
accordance with the transfer equation~(\ref{eq:transfer_eq}). From the ensemble
of packet trajectories, all relevant radiation field quantities can be
reconstructed.

With the increasing availability of computational resources, Monte Carlo
techniques have become a very popular and rewarding approach to radiative
transfer problems. Among the numerous applications of Monte Carlo radiative
transfer techniques in astrophysics are the calculation of mass-loss rates in
hot star winds \citep[see e.g.][]{Abbott1985, Lucy1993, Vink2000,
Sundqvist2010}, light curves and spectra of \snias{} \citep[see
e.g.][]{Mazzali1993,Lucy2005,Kasen2006,Sim2007,Kromer2009}, ionization structure
and synthetic spectra of photoionized nebulae \citep[e.g.][]{Ercolano2003},
mass-outflows in cataclysmic variables \citep[e.g.][]{Long2002, Noebauer2010} or active
galactic nuclei \citep[e.g.][]{Sim2005}.  Compared with classical ray-tracing
techniques, the Monte Carlo approach has certain advantages.  Most important,
from a physical viewpoint, is the ease with which complex scattering and
absorption processes can be incorporated.  Since all interactions with the
surrounding medium can be directly simulated during the propagation of the Monte
Carlo packets, even the most complex atomic processes can be included in the
radiative transfer calculation \cite[see e.g.][]{Lucy2002,Lucy2003, Lucy2005}.
These interactions are all simulated locally in the co-moving frame, making the
Monte Carlo algorithm entirely independent of the large-scale properties of
the simulation and readily applicable even to problems with arbitrarily
complex multi-dimensional geometries.

Apart from these physically motivated advantages, the Monte Carlo method also
brings computational benefits. As the propagation of one packet is independent
of the behaviour of all others, Monte Carlo radiative transfer calculations can
be easily parallelised and scale very well to large numbers of computational
cores. This is of great significance since the efficient use of high performance
computing facilities is an important consideration for the feasibility of
modelling complex astrophysical systems. 

Of course, Monte Carlo radiative transfer methods also have their drawbacks. The
accuracy and computational efficiency of the Monte Carlo approach are limited by
the number of packets that discretize the radiation field and by the number of
physical interactions they simulate. Consequently, whether Monte Carlo radiative
transfer methods are appropriate strongly depends on the specific problem under
consideration \citep[e.g.][]{Pincus2009}.  For example, in some applications, a
detailed radiative transfer treatment is not required and the dynamical
behaviour of the radiation field can be adequately addressed with approximate
methods which perform faster than the Monte Carlo approach
\citep[e.g.][]{Kuiper2010}.  Independently of the specific application, Monte
Carlo methods always introduce a certain level of statistical fluctuations in
the simulations. We shall return to the subject of minimising the influence of
this Monte Carlo noise in Sections~\ref{sec:estimator} and \ref{sec:noise}.

\subsection{Discretization}
\label{sec:discretization}

As mentioned above, in the Monte Carlo approach the radiation field is
discretized into packets. In early uses of the Monte Carlo machinery, e.g.\ in
\citet{Avery1968}, the number of photons described by a packet was held constant
throughout the simulation.  However, we follow \citet{Abbott1985} and instead
choose to discretize into packets of constant radiative energy.  At every
instant in the simulation, each packet represents a monochromatic parcel of
radiative energy -- i.e.\ a number of identical photons, all with a certain
frequency $\nu$. When packets interact with the fluid, both the number of
photons and the photon frequency associated with a packet can change but the
energy it carries in the local rest frame of the fluid remains fixed.  In
addition, the energy packets are indivisible -- i.e.\ processes may create or
destroy packets but never cause them to be split into multiple packets
\cite[see][]{Lucy2002,Lucy2003,Lucy2005}.  This approach is motivated by the
fact that total energy is conserved in interactions between the radiation field
and its surrounding medium but the number of photons generally is not. The
indivisible energy-packet method has been shown to be extremely powerful in
solving radiative equilibrium problems owing to the ease with which it can
ensure energy conservation \cite[see e.g.][]{Lucy1999}.  By extending the
method to include a net energy exchange between the radiation field and the
matter, we continue to exploit the efficiency of this approach together
with the properties of PPM to ensure global energy conservation in our
simulations. The use of indivisible packets also avoids computational
difficulties arising in cases where a fixed photon-number approach would require
that packets are split (e.g.\ in modelling fluorescence or recombination
cascades, where a physical process excited by a single photon leads to
the re-emission of many). With the indivisible packet method, these processes are
simulated with a probabilistic approach such that no packet splitting is
required but that all the cascade channels are correctly sampled when a
sufficiently large number of Monte Carlo packets are included
\cite[see][]{Lucy2002}.

With this discretization scheme, the Monte Carlo packets are naturally
well-suited to represent the mean intensity of the radiation field and its
temporal evolution. In addition, all derived radiation field characteristics can
be easily formulated and reconstructed from Monte Carlo estimators and fully
frequency-dependent opacities could be readily implemented. However, the
radiative flux is in general less accurately captured by the ensemble of energy
packets. We will discuss the implication of this in more detail when considering
statistical noise and the accurarcy of our approach in Section \ref{sec:noise}.

\subsection{Reference Frames}

An accurate and detailed description of the dynamical behaviour of the radiation
field requires that special relativistic effects are considered.  We have
therefore designed our Monte Carlo radiative transfer algorithm to
account for at least all first order special relativistic effects. In principle,
there are no obstacles to extending the method to higher order corrections.

For handling relativistic effects, we must clearly distinguish between two
reference frames. We define the spatial and temporal discretization of the
problem (i.e.\ our numerical grid) in the ``lab'' (or ``observer'') frame. The
initialisation and propagation of the Monte Carlo packets is also performed in
that frame.  However, the natural choice for the treatment of matter-radiation
interactions is the local rest frame of the fluid, which we will refer to as the
``co-moving'' frame. Henceforth, a subscript 0 will be used to identify
quantities that are defined in the local co-moving frame.

\subsection{Simplifications Adopted}
\label{sec:simplifications}

For the sake of clarity, in this section and below, we adopt some
simplifications that apply to our current implementation and the test problems
that we address in Sections 4 and 5.  In particular, we restrict ourselves
to one-dimensional problems, e.g.\ plane-parallel or spherically-symmetric
media. By arranging the problem setup such that the symmetry axis coincides with
the coordinate $z$-axis the radiation propagation direction can be specified by
the scalar
\begin{eqnarray}
  \mu = \bmath{n} \cdot \bmath{e}_z.
  \label{eq:direction_mu}
\end{eqnarray}
In spherically symmetric geometries, $\mu$ is measured with respect to the
radial direction. As a consequence of the symmetry properties of the problems we
consider, all radiation-hydrodynamical quantities only vary along one spatial
coordinate. Thus, the components of the radiation force in the other two
orthogonal directions ($G^1$ and $G^2$) vanish and will not be considered
further.

In addition to the geometry restrictions, we treat radiative transfer in a 
grey, i.e.\ frequency-independent, approximation. Scattering interactions between
the radiation field and the surrounding material are assumed to be coherent and
isotropic. We stress, however, that all these simplifications are not
necessary -- the approach can be readily generalized to multiple dimensions and
frequency-dependent opacities. Finally, we use an ideal gas equation of state to
relate fluid internal energy and thermodynamic pressure.

\subsection{Packet Initialisation}
\label{sec:initialisation}

At the beginning of a simulation, we need to generate an initial population of
Monte Carlo packets that describe the initial radiation field. This generation
process includes assigning each packet an initial position, direction and
frequency. All the steps involved in this process have to accommodate the
probabilistic nature of the Monte Carlo machinery. 

To initialize the population of Monte Carlo packets, an initial
condition for the radiation field has to be chosen. As an illustrative example, we assume
that the simulation is to be initialized in local thermodynamic equilibrium
(LTE). In this case, the radiation energy density in the co-moving frame, $E_0$,
follows the Stefan-Boltzmann law
\begin{eqnarray}
  E_{0} = a_{\mathrm{R}} T^4.
  \label{eq:stefan-boltzmann}
\end{eqnarray}
To correctly initialize the Monte Carlo packets, the total energy is first
transformed into the lab-frame and summed over the entire computational domain
\begin{eqnarray}
  \mathcal{E}_{\mathrm{tot}} = \sum_{i} \gamma^2 \left(1 +
  \frac{1}{3} \beta^2 \right) a_{\mathrm{R}} T_i^4 \Delta V_i .
  \label{eq:tot_energy}
\end{eqnarray}
Here, $\beta$ and $\gamma$ denote the usual parameters of special relativity
\begin{flalign}
  &\beta = \frac{u}{c},
  \label{eq:beta}\\
  &\gamma = \frac{1}{\sqrt{1 - \beta^2}},&
  \label{eq:gamma}
\end{flalign}
and the index $i$ runs over all grid cells.
The total energy $\mathcal{E}_{\mathrm{tot}}$ is then divided equally into a
chosen number of Monte Carlo packets (all packets are assigned the same
initial lab-frame energy), which are spread over the grid cells according to the
local radiative energy content.  The initial position of a packet within a grid
cell is chosen randomly.  In LTE, the radiation field is isotropic in the
co-moving frame, i.e.\ it has no angular dependence. Due to angle aberration
effects, this isotropy is lost during the transformation into the lab frame.
Consequently, the assignment of the initial propagation direction has to account
for the angular dependence of the radiation field in the lab frame. In the grey
approximation and under the restriction to one-dimensional problems, the LTE lab
frame specific intensity follows
\begin{eqnarray}
  I(\mu) = \frac{B(T)}{\gamma^4 (1 - \beta \mu)^4},
  \label{eq:specific_intensity_lf}
\end{eqnarray}
with $B(T)$ denoting the frequency-integrated Planck function 
\begin{eqnarray}
  B(T) = \frac{\sigma_{\mathrm{R}}}{\pi} T^4.
  \label{eq:planck-function}
\end{eqnarray}
This angular dependence can be translated into a probability density
\begin{eqnarray}
  \rho(\mu) = \frac{(1 - \beta^2)^3}{2 (1 + 1/3 \beta^2)} \frac{1}{(1 - \beta
  \mu)^4},
  \label{eq:mu_prob_density}
\end{eqnarray}
which can be sampled to give the relativistically correct directional
distribution of the initial Monte Carlo packets. In the classical limit ($\beta
\rightarrow 0$, $\gamma \rightarrow 1$), the density simplifies to
\begin{eqnarray}
  \rho(\mu) = \frac{1}{2},
  \label{eq:mu_prob_density_classical}
\end{eqnarray}
which can be easily sampled by the random number experiment
\begin{eqnarray}
  \mu = 2\xi - 1,
  \label{eq:mu_sample_classical}
\end{eqnarray}
where $\xi$ denotes a random number drawn uniformly
from the interval $[0,1]$.  In the more general case,
Equation~(\ref{eq:mu_prob_density}) must be sampled, leading to a more complex
expression for $\mu$ in terms of $\xi$, but the same principle applies.

In general, each packet also has to be assigned a photon frequency ensuring that
the packets represent the correct spectrum of the radiation field. In this work,
however, this step can be skipped since we currently adopt a grey-approximation.
A possible realisation of the more general sampling process can be found in
\citet{Lucy1999}.

\subsection{Sequence of Monte Carlo simulations}

\begin{figure}
  \resizebox{\hsize}{!}{\includegraphics{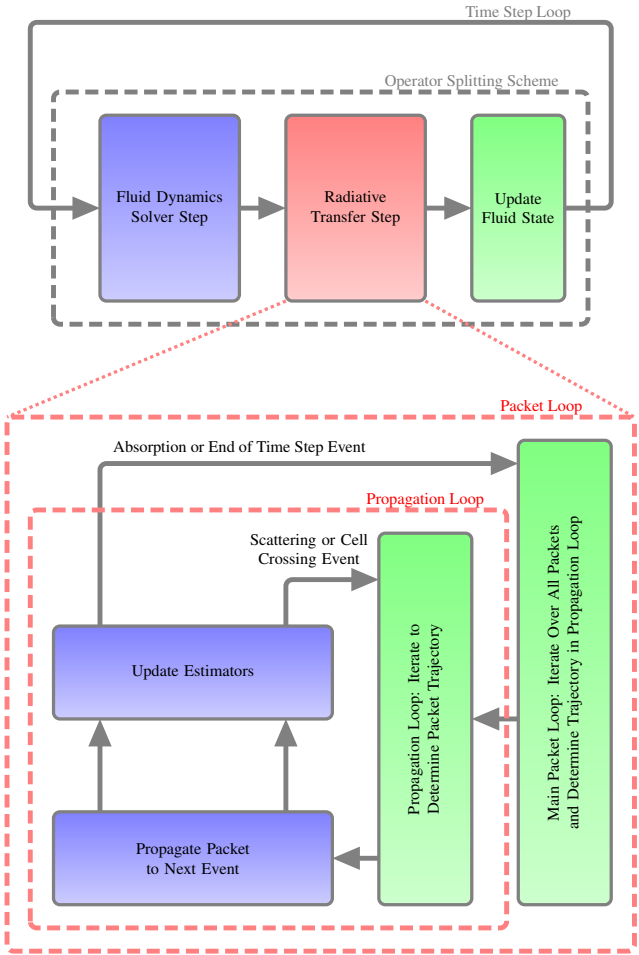}}
  \caption{Flow chart of the operator split algorithm and a detailed outline of
  the Monte Carlo radiative transfer step. During this process, the trajectories
  of all packets are determined and the radiation field characteristics are
  calculated by the Monte Carlo estimators.}
  \label{fig:chart}
\end{figure}

After their initialization at the beginning of the simulation, the Monte Carlo
packets are propagated through the medium.  During each time step, the packets
are able to move (see Section~\ref{sec:propagation}), some packets are
destroyed and others are created (see Section~\ref{sec:interaction}). At the end of
each time step, the properties of the currently active packets are stored so
that these packets can be reactivated at the start of the next time step, after
the fluid properties have been updated (see Section~\ref{sec:rad_influence}). A
graphical outline of the program flow is shown in Fig.~\ref{fig:chart}.

\subsection{Packet Propagation}
\label{sec:propagation}

In the Monte Carlo simulation, the packets propagate with the speed of light $c$
through the medium. To simulate the dynamical evolution of the radiation field,
the packets undergo interactions with the surrounding material as they
propagate. In the Monte Carlo method, the propagation and interactions are
treated stochastically.  In particular, the location of packet interactions are
determined by random number experiments -- the trajectory of a propagating
packet is terminated by an interaction with the surrounding medium when it
covered a path-length 
\begin{eqnarray}
  l = - \frac{\ln \xi}{\chi} ,
  \label{eq:interaction_length}
\end{eqnarray}
which follows from sampling the extinction law
\begin{eqnarray}
  I = I_0 \mathrm{e}^{-l \chi}.
  \label{eq:extinction_law}
\end{eqnarray}

In addition to simulating physical interactions in this stochastic manner, two
additional events that stem from the spatial and temporal discretization have to
be taken into account. During the propagation process, packets can cross grid
cell boundaries or reach the end of the current simulation time step. In the
case of cell crossings, the changing fluid properties have to be taken into
account, i.e.\ the interaction length $l$ has to be recalculated. Since the
Monte Carlo packets propagate with the speed of light, they will at maximum
travel a distance of $d = c \Delta t$ during a simulation cycle of duration
$\Delta t$. At this point the propagation of a packet is suspended and its
properties are stored in order to resume the propagation at the beginning of
the next time step. 

\subsection{Interaction Formalism}
\label{sec:interaction}

Since our Monte Carlo packets represent photon packets, their interactions
should model the physical interactions of photons with the surrounding
medium: scattering, absorption and emission processes.  When a packet
experiences a physical interaction (i.e.\ once it has propagated the path-length
given by Equation~\ref{eq:interaction_length}), we must first determine the
nature of the interaction.  If we include both scattering ($\sigma$) and
absorption ($\kappa$) contributions to the opacity
\begin{eqnarray}
  \chi = \sigma + \kappa,
  \label{eq:opacity_scatter_absorption}
\end{eqnarray}
the relative strength of these will determine the probability of the interaction
having been a scattering or absorption event.  In particular, we identify that a
packet scatters if 
\begin{eqnarray}
  \xi \le \frac{\sigma}{\sigma + \kappa}
  \label{eq:albedo}
\end{eqnarray}
is fulfilled. To treat a coherent scattering event, the packet is transformed
into the co-moving frame
\begin{eqnarray}
  \varepsilon_0 = \varepsilon \gamma (1 - \beta \mu)
  \label{eq:transform_energy_cmf}
\end{eqnarray}
where the packet energy is conserved and a new propagation direction $\mu_0'$ is
drawn isotropically (cf.\ Equation~\ref{eq:mu_sample_classical}). Afterwards,
the packet properties are transformed back into the lab frame 
\begin{flalign}
  &\varepsilon' = \varepsilon_0 \gamma (1 + \beta \mu_0'),&
  \label{eq:transform_energy_lf}\\
  &\mu' = \frac{\mu_0' + \beta}{1 + \beta \mu_0'},&
  \label{eq:transform_direction_lf}
\end{flalign}
to resume the propagation. For all other outcomes of the experiment
(Equation~\ref{eq:albedo}), an absorption event occurs, resulting in the
destruction of the packet. Consequently, this packet stops its propagation and
is no longer considered in the remaining simulation process. 

Destruction of packets by absorption interactions drains the energy content
of the radiation field.  However, radiation energy is also created due to emission
by the medium.  In the case of thermal emission, the radiative energy pool in a
grid cell of volume $\Delta V$ is increased by
\begin{eqnarray}
  \Delta \mathcal{E} = 4 \gamma \kappa \sigma_{\mathrm{R}} T^4 \Delta V \Delta t
  \label{eq:emission}
\end{eqnarray}
during a time step of size $\Delta t$ assuming constant temperature and opacity.
To represent this effect in the Monte Carlo calculation, a number of new packets
are created and launched during each cycle whose total energy is consistent with
this energy injection. The initial packet properties (e.g.\ direction of
propagation, which we assume to be isotropic in the co-moving frame) are
assigned by sampling appropriate probability distributions, in analogy to the
process of representing the initial radiation field at the onset of the
simulation (see Section~\ref{sec:initialisation}). However, these emitted
packets are not all injected into the simulation at the start of a time step but
at randomly determined times during the time step, thereby accounting for the
continuous energy injection by the thermal emission process.

In all interaction processes, momentum and/or
energy are transferred between the radiation field and the surrounding material.
Through their interaction behaviour, the packets account directly for the impact of
the momentum and energy flows on the radiation field. The complementing effect
on the surrounding fluid material is equally important (see
Equations~\ref{eq:rh_momentum_eq} and \ref{eq:rh_energy_eq}). These flow terms
can be reconstructed from the Monte Carlo simulation with the help of
\textit{volume-based Monte Carlo estimators} (see next Section,
\ref{sec:estimator}).

\subsection{Monte Carlo Estimators}
\label{sec:estimator}

The Monte Carlo packets were introduced to discretize the radiation field and,
during their propagation, simulate its interactions with the medium.  In
principle, all properties of the radiation field can be directly determined from
the instantaneous properties of the packets at any given time, e.g.\ from their
instantaneous distribution in real space, momentum space and frequency space at
the end of a time step. However, the accuracy with which radiation quantities
can be reconstructed is limited by Monte Carlo noise and it is advantageous to
consider how to extract maximum information from the Monte Carlo simulation.
\citet{Lucy1999} showed that radiation field quantities can be more accurately
determined by using volume-based Monte Carlo estimators rather than directly
counting properties of the packets. In this approach, the complete ensemble of
trajectories that packets traverse during a time step is used to reconstruct the
radiation field characteristics \citep[see ][]{Lucy1999, Lucy2003, Lucy2005}.
Thus, each packet contributes to the cell-averaged values of the radiation field
characteristics according to the fraction of the time that it spent in the cell.
Compared to reconstruction methods based on considering the final packet
distribution at the end of the propagation step, this cumulative approach
significantly reduces the Monte Carlo noise, which is in general the limiting
factor in the applicability of Monte Carlo methods.  To reconstruct the
cell-averaged radiation field energy density all packet trajectory segments that
cross the cell under consideration or lie within it are taken into account.  On
each of these trajectory segments, the packet contribution scales with its
energy $\varepsilon$. Every contribution is weighted according to the ratio of
the time $\delta t_i$ the packet spent on the trajectory element to the full
duration of the simulation time step $\Delta t$. Replacing the propagation time
with the path length $l_i = c \delta t_i$ of the individual segments and summing
up, gives the estimator
\begin{eqnarray}
  E = \frac{1}{\Delta V c \Delta t} \sum_i \varepsilon l_i
  \label{eq:estimator_energy}
\end{eqnarray}
for the radiation field energy density. Here, the summation runs over all
trajectory elements in one cell, implying that a packet may contribute several
times to the estimator. Using the fundamental relation between
the radiation field energy density and the mean intensity
\begin{eqnarray}
  J = \frac{c}{4 \pi} E,
  \label{eq:intensity_energy_relation}
\end{eqnarray}
a Monte Carlo estimator for the latter can be formulated
\citep[see][equation 12]{Lucy1999}
\begin{eqnarray}
  J = \frac{1}{4 \pi \Delta V \Delta t} \sum \varepsilon l.
  \label{eq:estimator_mean_intensity}
\end{eqnarray}
By restricting the estimator sum to contributions by packets propagating into a
certain directional bin $[\mu, \mu + \Delta \mu]$,
\begin{eqnarray}
  I(\mu) \Delta \mu = \frac{1}{2 \pi \Delta V \Delta t} \sum_{[\mu, \mu + \Delta
  \mu]}
  \varepsilon l.
  \label{eq:estimator_intensity}
\end{eqnarray}
the specific intensity can also be determined in a similar manner
\citep[see][equation 2]{Lucy2005}.  With this expression, estimators for all
radiation field characteristics that depend on the specific intensity or its
moments can be easily formulated.

For radiation-hydrodynamical problems, the radiation force components have to be
reconstructed from the Monte Carlo simulation to determine the energy
and momentum flow between the fluid and radiation field. The components of the
radiation force can be interpreted as the difference between absorbed and
emitted radiation energy and momentum respectively. We clearly separate the
contributions to the radiation force terms caused by scattering interactions
from the energy and the momentum transferred in pure absorption and emission
events. To identify the latter contribution, each packet trajectory is
considered to affect the cell-averaged absorbed energy, even if the packet did
not explicitly interact during the propagation cycle. Conceptually, we are
therefore counting all absorption events that \textit{could} have happened in
the simulation (weighted by their probability of occurring), rather than simply
counting the events that \textit{did} happen.  Thus, while propagating a
path-length $l$, a packet contributes with
\begin{eqnarray}
  \Delta E = \frac{l \kappa \varepsilon}{\Delta V}
  \label{eq:pure_absorption_contrib}
\end{eqnarray}
and with
\begin{eqnarray}
  \Delta p = \frac{\mu l \kappa \varepsilon}{c \Delta V}
  \label{eq:pure_absorption_momentum_contrib}
\end{eqnarray}
to the cell-averaged absorbed energy and momentum densities respectively.
The complementing energy and momentum injection due to
thermal emission is determined analytically for each cell. Accounting for the
frame transformation into the lab frame the injection rates can be formulated as 
\begin{flalign}
  &\dot E = 4 \gamma \kappa \sigma_{\mathrm{R}} T^4,&
  \label{eq:pure_absorption_emit_rate}\\
  &\dot p = \frac{4}{c} \beta \gamma \kappa \sigma_{\mathrm{R}} T^4.&
  \label{eq:pure_absorption_momentum_emit_rate}
\end{flalign}

In addition to determining the contribution of pure absorption and re-emission
events to the energy and momentum transfer terms, scattering interactions have
to be incorporated. We consider scattering interactions formally as a
combination of an absorption event, followed immediately by the coherent and
isotropic re-emission of a Monte Carlo packet. This interpretation allows us to
reconstruct the scattering contribution analogously to
Equations~(\ref{eq:pure_absorption_contrib}) and
(\ref{eq:pure_absorption_momentum_contrib}). In general, however, formulating an
analytic expression to quantify the re-emission part is non-trivial. This
difficulty can be circumvented by exploiting the fact that Monte Carlo packets
are not destroyed in scattering interactions. Thus, the packets can simulate
both the absorption and re-emission parts of the scattering events at the same
time. By additionally assigning each packet a set of hypothetical
``post-scatter'' properties, i.e.\ drawing a propagation direction $\mu_0'$ that
translates into an energy $\varepsilon'$  and a direction $\mu'$ in the lab
frame according to the transformation rules~(\ref{eq:transform_energy_lf}) and
(\ref{eq:transform_direction_lf}), each packet trajectory $l$ contributes to
cell averaged energy and momentum transfer according to
\begin{flalign}
  &\Delta E = \frac{\sigma l}{\Delta V} (\varepsilon^i - \varepsilon^f),&
  \label{eq:pure_scattering_contrib}\\
  &\Delta p = \frac{\sigma l}{c \Delta V} (\mu^i \varepsilon^i -  \mu^f
  \varepsilon^f).&
  \label{eq:pure_scattering_momentum_contrib}
\end{flalign}
Here, the superscripts $i$ and $f$ denote the current packet properties during
the propagation and the formal post-scattering state, respectively.

Finally, volume-based Monte Carlo estimators for the radiative force terms $G^0$
and $G^3$ are obtained by gathering all contributions of scattering and pure
absorption/emission events
\begin{flalign}
  &G^0 = \frac{1}{\Delta V c \Delta t} \sum l (\chi \varepsilon^i - \sigma
  \varepsilon^f) - \frac{4}{c}\gamma \kappa
  \sigma_{\mathrm{R}} T^4,&
  \label{eq:estimator_radf0}\\
  &G^3 = \frac{1}{\Delta V c \Delta t} \sum l (\chi \mu^i \varepsilon^i -
  \sigma \mu^f \varepsilon^f) - \frac{4}{c} \beta \gamma
  \kappa \sigma_{\mathrm{R}} T^4.&
  \label{eq:estimator_radf3}
\end{flalign}
Once transformed into the local co-moving frame, these estimators take the
expected form for the corresponding radiation field characteristics measured in
this frame.  In particular, the estimators (\ref{eq:estimator_radf0}) and
(\ref{eq:estimator_radf3}) become
\begin{flalign}
  &G^0_0 = \frac{1}{\Delta V_0 c \Delta t_0} \sum \kappa_0 l_0 \varepsilon_0 -
  \frac{4 \pi}{c} \kappa_0 B(T),&
  \label{eq:estimator_radf0_cmf}\\
  &G^3_0 = \frac{1}{\Delta V_0 c \Delta t_0} \sum \chi_0 l_0 \mu_0
  \varepsilon_0,&
  \label{eq:estimator_radf3_cmf}
\end{flalign}
after transformation into the local co-moving frame\footnote{For detailed
  derivations of the transformation laws for the radiation field
  characteristics, see e.g.\ \cite{Mihalas1984}.} using
\begin{flalign}
  &G^0_0 = \gamma (G^0 - \beta G^3),&
  \label{eq:transformation2cmf_radf0}\\
  &G^3_0 = \gamma (G^3 - \beta G^0).&
  \label{eq:transformation2cmf_radf3}
\end{flalign}
These co-moving frame estimators for $G^0_0$ and $G^3_0$ reproduce exactly the
radiative energy and momentum sources in this frame, which can be formulated as
\begin{flalign}
  &G^0_0 = \frac{4 \pi}{c} \kappa_{0} (J_{0} - B(T)),&
  \label{eq:radf0_analytic_cmf}\\
  &G^3_0 = \frac{\chi_0}{c} F_0.&
  \label{eq:radf3_analytic_cmf}
\end{flalign}
To obtain the same transformation behaviour for the Monte Carlo estimators in
spherically-symmetric geometries, the direction scalar has to be averaged over
the trajectory segment.

\subsection{Monte Carlo Noise}
\label{sec:noise}

Due to the stochastic nature of the Monte Carlo technique, all associated
quantities are subject to statistical fluctuations, limiting the accuracy of the
radiative transfer calculation. The obvious approach to reduce the Monte Carlo
noise by increasing the number of active packets is limited by computational
efficiency considerations. In this context, the concept of volume-based Monte
Carlo estimators, as described in the previous section is vital since it allows
us to extract the maximum amount of information from the propagation behaviour
of a given number of radiation packets. The reduction of the statistical
fluctuations in the reconstructed quantities is significant but still limited.
In particular, the noise level in the energy-momentum transfer and thus the
accuracy with which the radiation-hydrodynamical coupling is captured, varies
with the state of the radiation field, due to our discretization scheme and to
the form of the radiation force components in the co-moving frame.
Reconstructing the energy transfer from a difference of two terms as in Equation
(\ref{eq:radf0_analytic_cmf}), yields accurate results if the contributions are
clearly separable. However, if the mean intensity deviates from the equilibrium
configuration (i.e. $B(T)$) only on the level of the Monte Carlo noise, the
resulting heating/cooling term will be obscured by statistical fluctuations.
Similarly, the momentum transfer is only accurately captured if the radiative
flux is well-resolved (cf.\ Equation \ref{eq:radf3_analytic_cmf}). To achieve
high accuracy in the reconstruction via the estimator formalism, the packet
contributions to the estimators have to be much smaller than the reconstructed
quantity. This is the case for the mean intensity, since our discretization
scheme (indivisible energy packets) is most naturally suited to well-represent
this radiation field characteristic. In the reconstruction of the radiative
flux, however, the packet contributions to the estimator are weighted with the
propagation direction.  As a consequence, the radiative flux is only
well-resolved in our approach if $F_0 \lesssim 4 \pi J_0$, but obscured by Monte
Carlo noise for $F_0 \ll 4 \pi J_0$, i.e.\ for nearly isotropic radiation field
configurations.

In summary, our Monte Carlo approach is ideal for describing the mean
intensity and our estimators will accurately capture the radiation forces when
significant deviations from LTE or from isotropy are present. However, in
cases where the radiation field is nearly isotropic or close to the LTE
configuration, the radiation hydrodynamical effects, as described in the
radiation force terms, may be subject to significant statistical fluctuations.
In order to achieve a further suppression of the noise in this regime, we have found
it useful to smooth the reconstructed radiation field quantities over
neighbouring grid cells. This additional noise reduction method will be of
importance when addressing \snia{} ejecta (see Section \ref{sec:w7}).

\subsection{Dynamical Influence of the Radiation Field}
\label{sec:rad_influence}

With the numerical methods described in the previous sections, the temporal
evolution of the radiation field is determined in a Monte Carlo step as part of
the operator splitting approach to radiation hydrodynamics. After solving the
fluid dynamical evolution using PPM and simulating the radiative transfer in the
Monte Carlo simulation, the calculation of the new radiation-hydrodynamical
state ends with the inclusion of the radiative influence on the dynamics
specified by $G^0$ and $G^3$.  These terms are reconstructed from the Monte
Carlo simulation using the estimators defined in
Equations~(\ref{eq:estimator_radf0}) and (\ref{eq:estimator_radf3}) and describe
the energy and momentum transfer onto the surrounding material. Consequently the
fluid energy and momentum are updated according to 
\begin{eqnarray}
  \rho\frac{\partial}{\partial t} u = G^3,
  \label{eq:splitting_uupdate}\\
  \rho\frac{\partial}{\partial t} e = c G^0,
  \label{eq:splitting_eupdate}
\end{eqnarray}
as the second and last step of the Godunov splitting approach. This concludes
the calculation of one time step and a new simulation cycle is entered by
stepping through the segments of the splitting scheme again, starting by solving
the fluid dynamics for the next time step (see Fig.~\ref{fig:chart}).

\section{Testing}
\label{sec:testing}

The methods described in the previous section have been implemented into a
numerical code, hereafter referred to as \textsc{mcrh}. Currently, the program
is able to calculate the temporal evolution of radiating fluids in an Eulerian
or Lagrangian reference frame, assuming grey radiative transfer and
one-dimensional geometries (see Section~\ref{sec:simplifications}).  Before
using the code to model radiative flows in astrophysical environments or
implementing more complex geometries and opacities, the computational
feasibility and physical accuracy of our approach and its implementation has to
be verified. PPM is commonly used to simulate fluid flows in astrophysical
applications, e.g.\ in Type Ia and II supernova explosions \citep[e.g.
][]{Roepke2005a, Janka1996} or in relativistic jets \citep[e.g.
][]{Marti1997} and has already been tested extensively. Consequently, the main
focus of this discussion lies in testing the radiative transfer and the
radiation-matter coupling components of our approach.

We start this verification process by first testing the propagation behaviour of
the Monte Carlo packets. In the next stage, the interaction mechanism is checked
by examining the equilibration behaviour in a series of toy simulations.
Finally, we turn towards radiative shocks, the standard test problem for
radiation hydrodynamics. In calculating sub- and supercritical shocks, the
workings of both the radiation-matter coupling and our complete radiation
hydrodynamics method can be verified. 

\subsection{Random Walk}

As presented in Section~\ref{sec:propagation}, the Monte Carlo packets undergo a
multitude of physical interactions and numerical events on their propagation
trajectories. Since this behaviour is essential for accurately simulating the
temporal evolution of the radiation field, our first test calculation aims to
verify the packet propagation process. For this purpose, we follow
\citet{Harries2011} and consider a purely scattering medium in the optically
thick limit. The entire radiation field energy is initially concentrated in the
central cell of the computational domain (which has length $L$), mimicking a
$\delta$-function. In this problem, the initial radiation peak disperses
according to the diffusion equation
\begin{eqnarray}
  \frac{\mathrm{d} E}{\mathrm{d} t} = - D \frac{\mathrm{d}^2
  E}{\mathrm{d} z^2},
  \label{eq:diffusion_equation}
\end{eqnarray}
yielding
\begin{eqnarray}
  E(z, t) = E(t_0) \frac{L}{\sqrt{4 \pi D t}} \exp \left( - \frac{z^2}{4 D t}
  \right).
  \label{eq:heat_kernel}
\end{eqnarray}
With interactions restricted to scattering only, the diffusion coefficient is
given by 
\begin{eqnarray}
  D = \frac{c}{3 \sigma}.
  \label{eq:diffusion_coefficient}
\end{eqnarray}
In a Monte Carlo simulation of this problem, the packets are expected to perform
a random walk due to the large scattering optical depth. Consequently, the
temporal evolution of the radiative energy constitutes a suitable test to verify
the correct propagation behaviour of the Monte Carlo packets. 

\begin{figure}
  \resizebox{\hsize}{!}{\includegraphics{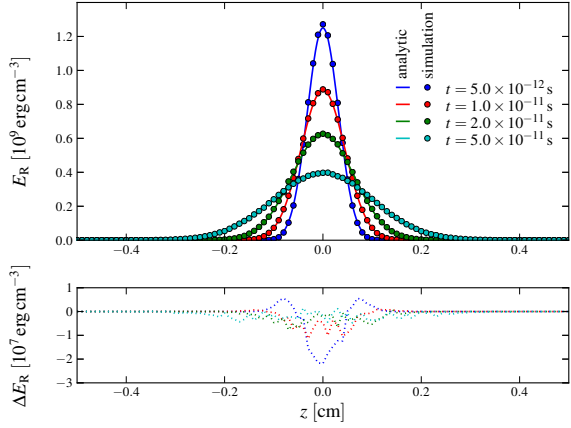}}
  \caption{Comparison between the theoretically expected diffusion behaviour
  (solid lines) of a centrally peaked radiation field in an optically thick,
  pure scattering medium and the corresponding Monte Carlo simulation
  (circles). The lower panel shows the absolute difference
  between the theoretocally expected ($E_{\mathrm{th}}$) and
  simulated ($E_{\mathrm{sim}}$) energy densities,
  $\Delta E_{\mathrm{R}} = E_{\mathrm{th}} - E_{\mathrm{sim}}$.}
  \label{fig:diffusion}
\end{figure}

Following \citet{Harries2011}, the test simulation is performed in a planar
symmetric box of length $L = 1 \, \mathrm{cm}$ that is divided into 101 equally
spaced cells. In the central cell, an initial radiation field energy density of
$E = 10^{10} \, \mathrm{erg} \, \mathrm{cm^{-3}}$ is deposited and discretized
by $10^5$ Monte Carlo packets. Fig.~\ref{fig:diffusion} shows the temporal
evolution of the radiative energy density in our Monte Carlo simulation. The
results are compared with the theoretically expected behaviour set by
Equation~(\ref{eq:heat_kernel}). The excellent agreement between the simulation
and the theoretical predictions demonstrates the accurate operation of the basic
Monte Carlo processes driving the packet propagation.

\subsection{Equilibration Behaviour}
\label{sec:equilibration}

Equally important as the correct propagation behaviour is the accurate
calculation of the momentum and energy transfer between the radiation field and
the surrounding material. We verify this part of our scheme by examining the
relaxation behaviour towards equilibrium in a series of toy simulations.
Following \citet{Turner2001} and \citet{Harries2011}, we consider a radiating
fluid initially far from equilibrium. In these particular tests, only absorption
and thermal emission events occur in the medium and all influences of the
thermodynamic and radiation pressure can be neglected. Therefore, the dynamical
behaviour of the radiating fluid can be expressed in terms of the radiation
field energy density $E_{\mathrm{R}}$ and the internal energy density
$E_{\mathrm{G}}$, which follow \citep[see][equation 23]{Harries2011}
\begin{eqnarray}
  \frac{\partial}{\partial t} E_{\mathrm{G}} = c \kappa E_{\mathrm{R}} - 4
  \kappa \sigma_{\mathrm{R}} T_{\mathrm{G}}^4(E_{\mathrm{G}})
  \label{eq:change_intE}
\end{eqnarray}
and
\begin{eqnarray}
  \frac{\partial}{\partial t} E_{\mathrm{R}} = - \frac{\partial}{\partial t}
  E_{\mathrm{G}}.
  \label{eq:change_radE}
\end{eqnarray}

\begin{figure}
  \resizebox{\hsize}{!}{\includegraphics{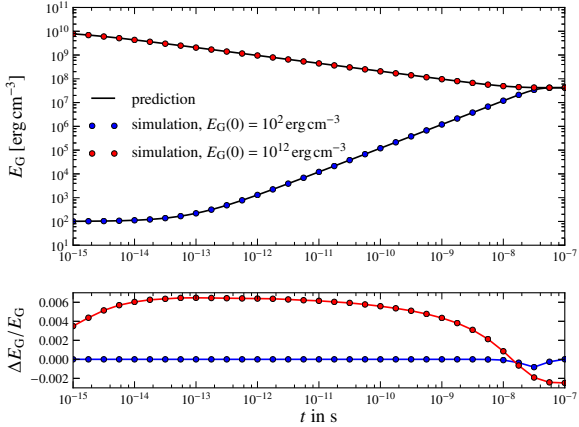}}
  \caption{Results of the equilibration simulations described in
    Section~\ref{sec:equilibration}.  The blue symbols show the results of the
    Monte Carlo simulation with an initial internal energy density of
    $E_{\mathrm{G}}(0) = 10^{2} \, \mathrm{erg \, cm^{-3}}$ for a series of
    selected snapshots. The corresponding results obtained for the initial energy density
    of $E_{\mathrm{G}}(0) = 10^{10} \, \mathrm{erg \, cm^{-3}}$ are shown with
    red circles. The temporal evolution of the internal energy density as
    predicted by Equation~(\ref{eq:change_intE}) for both cases is indicated by
    the solid black lines. The initial radiation field energy density is set to
    $E_{\mathrm{R}}(0) = 10^{12}\, \mathrm{erg\, cm^{-3}}$ in both simulations.
    In the lower panel, the relative difference between the theoretically
    expected ($E_{\mathrm{th}}$) and the simulated ($E_{\mathrm{sim}}$)
    evolution of the internal energy density, $(E_{\mathrm{th}} -
    E_{\mathrm{sim}}) / E_{\mathrm{th}}$, is shown for both calculations.}
  \label{fig:equil}
\end{figure}

As a first test, we follow the relaxation behaviour of the radiation field and
the internal gas energy in a one-dimensional plane-parallel box of length $L = 4
\, \mathrm{cm}$, which is discretized on 4 cells\footnote{The finite-volume
hydrodynamical solver based on PPM requires a domain size of at least 4 cells
due to the parabolic reconstruction and the discontinuity detection
\citep{Colella1984}.}. For this, we adopt the parameters $\rho = 10^{-7} \,
\mathrm{g \, cm}^{-3}$, $E_{\mathrm{R}}(0) = 10^{12} \, \mathrm{erg \,
cm}^{-3}$, $\kappa = 4 \times 10^{-8} \, \mathrm{cm}^{-1}$, $u = 0$ and the
adiabatic index $\gamma = 5/3$.  For these parameters, the equilibrium value for
the internal energy of the fluid is $E_{\mathrm{G}}^{\mathrm{eq}} =
4.2\times10^7 \, \mathrm{erg \, cm}^{-3}$.  We perform the relaxation test
twice, first with the initial internal energy of the fluid set to
$E_{\mathrm{G}}(0) = 10^{10} \, \mathrm{erg \, cm}^{-3}$ and then with
$E_{\mathrm{G}}(0) = 10^{2} \, \mathrm{erg \, cm}^{-3}$.  To 
predict the evolution of the systems, only Equation~(\ref{eq:change_intE}) has to
be considered, since the radiation field energy can be assumed to be constant in
this configuration.  The theoretical internal energy evolution, obtained from
numerical integration of Equation~(\ref{eq:change_intE}) for both setups is
compared with the results of the corresponding Monte Carlo simulations in
Fig.~\ref{fig:equil}. As illustrated in detail in the lower panel, the agreement
is excellent. Only a systematic deviation persists, which is caused by
the finite time resolution in the simulation. In the operator-split approach,
the gas temperature is assumed to be constant during the radiation transfer
cycle. Thus, the emissivity and in turn the amount of emitted energy is slightly
overestimated.

\begin{figure}
  \resizebox{\hsize}{!}{\includegraphics{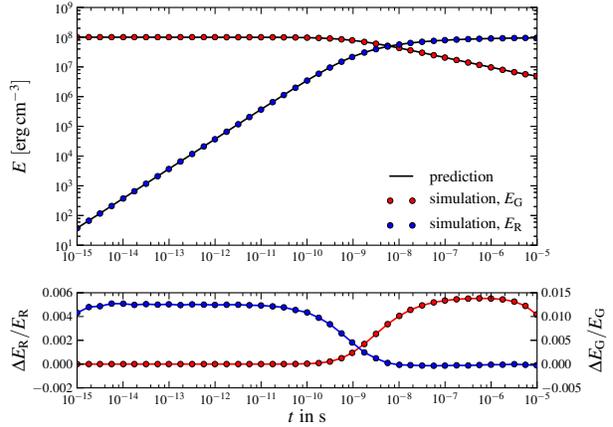}}
  \caption{Equilibration calculation with an initially absent radiation field
    and a starting internal energy density of $E_{\mathrm{G}} = 10^8 \,
    \mathrm{erg\, cm^{-3}}$. For a series of selected time snapshots,  the theoretical
    evolution of this system, obtained by numerically integrating Equations
    (\ref{eq:change_radE}) and (\ref{eq:change_intE}) is displayed as black
    lines. The corresponding results form our Monte Carlo simulation are
    depicted as blue and red circles for the radiative and internal energy
    density, respectively. As in Fig. \ref{fig:equil}, the relative
    difference between the theoretically expected and the simulated energy
    evolution is shown in the lower panel.}
  \label{fig:equil_zero}
\end{figure}

In a second test calculation, the initial radiation field energy content was set
to zero and the internal energy density to $E_{\mathrm{G}}(0) = 10^{8} \,
\mathrm{erg \, cm}^{-3}$. The remaining simulation parameters are adopted from
the previous equilibration setup. Over time, the fluid will drive the radiation
field towards equilibrium by the emission of thermal photons.  To theoretically
predict the evolution of the radiation-matter state, the coupled differential
Equations~(\ref{eq:change_radE}) and (\ref{eq:change_intE}) can be solved
simultaneously by numerical integration.  In contrast to the previous
simulations, neither the radiation temperature, nor the gas temperature can be
assumed to be constant. In Fig.~\ref{fig:equil_zero}, the results of our Monte
Carlo simulation are tested against the theoretically predicted evolution.
Again, only systematic deviations, which are caused by the finite duration of
the simulation time steps, persist on the per cent level.

The agreement between our test calculations and the theoretically predicted
behaviour for the relaxation towards equilibrium is excellent as illustrated in
Figs.~\ref{fig:equil} and~\ref{fig:equil_zero}, indicating that our
numerical methods describing the matter-radiation interactions and their
implementation are accurate and work correctly.

\subsection{Radiative Shocks}

The calculations presented above probed the propagation of the radiation field,
the basic interaction machinery and the determination of heating and cooling
terms, neglecting any effects of radiation and thermodynamic pressure and the
fluid motion. In our final, most challenging test, we verify that the
interaction processes and our approach as a whole yield the correct
radiation-hydrodynamical coupling. For this purpose we examine \textit{sub- and
supercritical radiative shocks}. The seminal works describing this type of
shocks analytically date back to \cite{Zeldovich1957} and
\cite{Raizer1957}\footnote{The corresponding translations to English can be
  found in \cite{Zeldovich1957a} and \cite{Raizer1957a}.} and to
  \cite{Marshak1958}. The results of the two former studies are summarized in
  detail by \cite{Zeldovich1967}.

Radiative shocks exhibit a structure and dynamical behaviour that is very
different from their purely hydrodynamical counterparts, due to the presence of
a radiative precursor. The pre-shock material is heated by the flow of radiative
energy through the shock front.  Depending on the strength of the shock and in
turn on the amount of heating of the pre-shock material, two classes of
radiative shocks can be distinguished.  Following the discussion of
\citet{Mihalas1984}, we denote the gas temperature behind the shock front with
$T_2$ and immediately in front of it with $T_{-}$.  In the case of $T_{-}$ being
much lower than $T_2$, due to the small amount of pre-heating, a
\textit{sub-critical shock} is encountered. With increasing shock strength,
equivalent to the shock front moving faster, the radiative precursor heats the
material more and more until it reaches the critical configuration of $T_{-} =
T_2$. At this point, a further increase in the shock strength only results in a
deeper penetration of the radiation precursor.  Consequently, $T_{-}$ never
surpasses the temperature behind the front in such \textit{supercritical
shocks}.

The structure of radiative shocks has already been determined in various
numerical studies, including works by \citet{Heaslet1963} and
\citet{Sincell1999}. Following the proposal of \citet{Ensman1994}, these shocks
have been used as a common test problem to verify the accuracy of
numerical approaches to radiation hydrodynamics. In particular, the realisation
of the matter-radiation coupling in the \textsc{zeus}-code has been tested
extensively using radiative shocks \citep{Turner2001, Hayes2003, Hayes2006}. For
comparison to our Monte Carlo radiation-hydrodynamical simulations, we will
use results calculated with the latest version of this code, \textsc{zeus-mp2}
\citep{Hayes2006}. 

\begin{figure}
  \resizebox{\hsize}{!}{\includegraphics{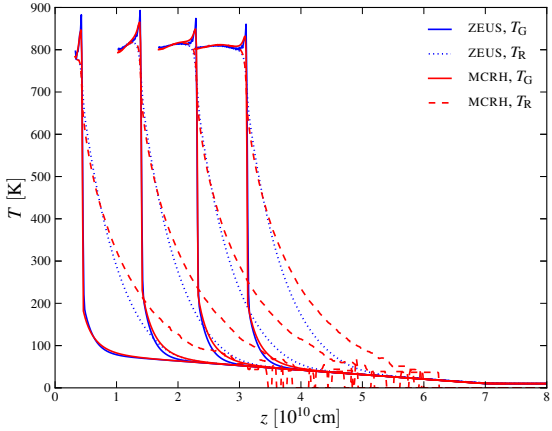}}
  \caption{Temperature profiles of a sub-critical radiative shock calculated
  with our Monte Carlo radiation hydrodynamics (red lines) program and the
  \textsc{zeus-mp2} code (blue lines). The gas temperature is depicted with solid
  lines, while the radiation temperature is shown as dashed and dotted lines.
  Profiles are shown for $t = 5.4 \times 10^3$, $1.7 \times 10^4$, $2.8 \times
  10^4$ and $3.8 \times 10^4 \, \mathrm{s}$ and are plotted in the rest frame of
  the un-shocked material. Note the Monte Carlo noise in the radiation
  temperature ahead of the shock. Here, the radiation field is only represented by
  a small number of packets.}
  \label{fig:subcritical}
\end{figure}

\begin{figure}
  \resizebox{\hsize}{!}{\includegraphics{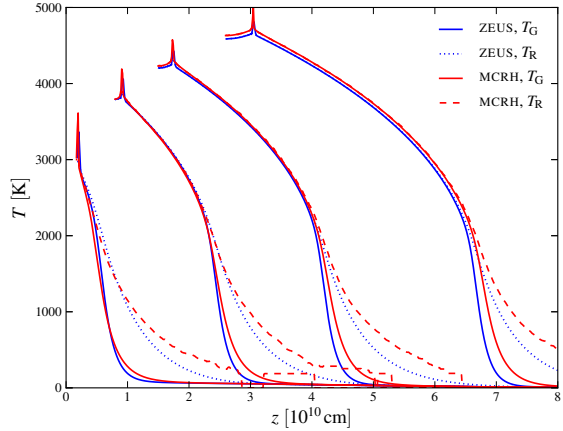}}
  \caption{Analogous to Fig.~\ref{fig:subcritical} but now displaying the
  corresponding calculation of the supercritical shock. The results from the
  Monte Carlo simulation are again shown in red and the ones obtained with
  \textsc{zeus-mp2} in blue. Gas and radiation temperature are indicated with solid and
  dashed/dotted lines respectively and shown with respect to the un-shocked
  material. Temperature profiles are presented for $t = 8.6 \times 10^2$, $4.0
  \times 10^3$, $7.5 \times 10^3$ and $1.3 \times 10^4 \, \mathrm{s}$.}
  \label{fig:supercritical}
\end{figure}

For the calculation of sub- and supercritical radiative shocks we choose the
properties of the radiating fluid according to \citet{Hayes2006} which in turn
were motivated by the simulations performed by \citet{Ensman1994}. In particular,
a one-dimensional plane-parallel domain of length $7 \times 10^{10} \,
\mathrm{cm}$ and a fluid with initial density of $7.78 \times 10^{-8} \,
\mathrm{g \, cm}^{-3}$ is considered. The temperature is set to $85 \,
\mathrm{K} $ at the inner, reflecting boundary and decreased linearly to $10 \,
\mathrm{K} $ at the outer open boundary of the domain.  We adopt this profile to
facilitate comparison with the calculations performed by \citet{Ensman1994}, in
which a temperature gradient had to be imposed for reasons of numerical
stability. Throughout the domain, a uniform grey absorption opacity of $3.1
\times 10^{-10} \, \mathrm{cm}^{-1}$ is chosen, resulting in a photon mean free
path that is roughly 20 times shorter than the extent of the computational box.
The shock is created by driving the material into the reflecting inner boundary
with a bulk velocity of $-6 \, \mathrm{km \, s}^{-1}$ for the sub-critical case
and with $-20 \, \mathrm{km \, s}^{-1}$ for the supercritical calculation,
respectively. We determine the evolution of the resulting radiative shocks with
our Monte Carlo program and as a reference with the \textsc{zeus-mp2} code.
Figs.~\ref{fig:subcritical} and \ref{fig:supercritical} show comparisons
between the gas and radiation temperature structures calculated with both codes
for the sub- and the supercritical shocks respectively. To be compatible with
\citet{Ensman1994} we display the shock structure in the rest frame coordinates
of the un-shocked material \citep{Hayes2003,Hayes2006}.

As shown in Figs.~\ref{fig:subcritical} and \ref{fig:supercritical}, the
radiative shocks calculated with our \textsc{mcrh} code exhibit the expected
overall structure and characteristic features. In the subcritical case, a weak
radiative precursor leads to mild heating of the pre-shock material. The
effect of the precursor is much more prominent in the second case of the
supercritical shock. Here, the sharp shock front is washed out by the strong
heating of the upstream material by the radiative flux penetrating the pre-shock
domain. As predicted for supercritical shocks, the material immediately ahead
of the shock is heated significantly, reaching the same temperature as behind
the front. Overall, our numerical results agree very well with the
\textsc{zeus-mp2} calculations, especially with respect to the location of the
shock and the gas temperature profiles. However, our Monte Carlo simulations
predict a stronger and deeper penetration of the unshocked material by the radiation
field. This results in increased heating of the material and differences in the
radiation temperature profiles with respect to the \textsc{zeus-mp2} results. These
differences are most likely caused by the simplifications implemented
in the \textsc{zeus-mp2} code, which treats the radiative flux in the
diffusion approximation, augmented by the introduction of a flux limiter
\citep[see][equation 11]{Hayes2006}. The influence of this simplified scheme
on the structure of radiative shocks has already been studied by
\citet{Ensman1994}. Calculating the evolution of a supercritical shock on the one
hand by relying on the diffusion approximation and on the other hand by solving
the radiation-hydrodynamical equations without any simplifications directly,
revealed the same behaviour of the radiative precursor \citep[see][fig.\
15]{Ensman1994}. Despite its statistical nature, the Monte Carlo approach
provides a direct solution to the transfer equation without introducing any
physical simplifications. For these reasons we believe that the differences are
due to the employed methods and not a shortcoming of our approach to
radiation hydrodynamics.

The test calculations presented in this section have been carried out
without using the smoothing capability of our program (see Section
\ref{sec:noise}). In the simulations determining the structure of radiative
shocks, the noise suppression of the volume-based estimator approach was
sufficient to accurately resolve the heating effects in the precursor regime
with large deviations from LTE. For example, in our simulation of the
subcritical shock a maximum of $6 \times 10^5$ packets were active at a time,
simulating the evolution of the radiation field. However, as anticipated
in Sections \ref{sec:discretization} and \ref{sec:noise}, in the regions
around the shock front, where the radiation field is close to LTE, the
radiation force components are subject to non-negligible statistical fluctuations.
But this noise component is effectively suppressed on the characteristic
hydrodynamical time scales, due to a high packet recycling rate. Typically,
the entire packet ensemble is re-populated multiple times during a radiative
transfer step, causing a smooth temperature profile even in the near-LTE
regions.

With the calculations presented in this section, the different individual
mechanisms of our method have been successfully tested and the correct
radiation-hydrodynamical behaviour of the approach as a whole has been
demonstrated without applying any smoothing. A first application to an
astrophysical environment is presented in the next Section.

\section{Application}
\label{sec:application}

Our framework has been specifically designed to address radiation-hydrodynamical
problems in astrophysical systems with large outflows such as supernova (SN)
explosions. To verify our approach for problems of astrophysical
interest, we consider the homologous expansion phase of \snia{} ejecta.

In the expelled material of \snias{}, radioactive decays emit $\gamma$-rays,
which interact strongly with the ejecta. This coupling affects the expansion
dynamics leading to changes in the density and velocity profile that could
influence the observable display. The question of whether this
radiation-hydrodynamical effect is important for theoretical determinations of
SN light curves and spectra has already been addressed by \citet{Pinto2000}
using a series of analytical estimates. This study concluded, that the radiative
influence on the expansion of the ejecta is marginal and does not cause
significant changes in the light curve. These findings were verified in the
study of \citet{Woosley2007}, which involved detailed radiation-hydrodynamical
calculations with the \textsc{stella} code \citep{Blinnikov2006}. Changes on a
10 per cent level in the density stratification of the ejecta were identified
when including the radiation-matter coupling. Here, we re-address this problem
for the purpose of testing our radiation-hydrodynamical approach in
astrophysical environments.

\subsection{Code Extensions}
\label{sec:extensions}

For the application to supernova ejecta, our program is extended to account for
the energy release accompanying radioactive decays. The determination of the
energy injection into the radiation field involves tracking the abundances of
\elem{Ni}{56} and its daughter nucleus \elem{Co}{56}. Both radio-nuclides are
proton-rich and decay through electron-capture reactions\footnote{Note that the \elem{Co}{56} decay proceeds via positron
emission in about 20 per cent of cases. As in \cite{Lucy2005}, we assume
instantaneous local annihilation of the positrons yielding a pair of
$\gamma$-rays. We neglect the kinetic energy of the positrons.}  down to stable
\elem{Fe}{56}. These decay reactions occur on the characteristic $e$-folding
time scales of $t_{\mathrm{Ni}} = 8.80 \, \mathrm{d}$ for $\melem{Ni}{56}
\rightarrow \melem{Co}{56}$ and $t_{\mathrm{Co}} = 113.7 \, \mathrm{d}$ for
$\melem{Co}{56} \rightarrow \melem{Fe}{56}$ and are accompanied by a cascade of
$\gamma$-ray emissions. The $\gamma$-photons interact with the surrounding
material through Compton-scattering, production of $e^+ e^-$ pairs and the
photoelectric effect. As a result of these interaction mechanisms, the
$\gamma$-radiation heats the surrounding material and the energy is re-radiated
as
quasi-thermal emission. Assuming instant thermalisation we simplify the
$\gamma$-interaction processes by relying only on one effective absorptive
opacity \citep{Sutherland1984,Swartz1995}. We model the net effect of injecting
energy into the thermal radiation field by a grey Monte Carlo transport step.
For each simulation time step, the number of radioactive decays is determined
and an adequate number of Monte Carlo packets representing the emitted
$\gamma$-energy is created. For this purpose we follow \citet{Lucy2005} and
integrate over the $\gamma$-spectrum of the decay reactions \citep[see
][table 1]{Ambwani1988}, obtaining a total energy release in the form of
$\gamma$-radiation of $E_{\mathrm{Ni}} = 1.728 \, \mathrm{MeV}$ and
$E_{\mathrm{Co}} = 3.566 \, \mathrm{MeV}$ per decaying \elem{Ni}{56} and
\elem{Co}{56} nucleus respectively. The $\gamma$-packets propagate in the same
manner as the Monte Carlo packets describing the thermal radiation field, but
the interactions with the ejecta material are described by different opacities.
Each $\gamma$-packet that undergoes an absorption is automatically transformed
into a thermal radiation packet, which is then treated according to the methods
laid out in Section~\ref{sec:propagation}.

\subsection{Toy Simulations}
\label{sec:lucymodel}

To verify the correct operation and the validity of the adopted simplified
treatment of the radioactive energy injection, we present the results of a
simple toy simulation following \citet{Lucy2005}. In his study, Lucy developed a
Monte Carlo radiative transfer method to determine spectra and light curves in
\snias{}. In the verification process, Lucy performed a grey radiative transfer
simulation for a simplified ejecta structure under the assumption of homologous
expansion. A comparison of the bolometric light curve determined in the Monte
Carlo simulation with the results of a moments-equation solution technique
\citep{Castor1972} yielded excellent agreement. Using the results of
\citet{Lucy2005} as a reference, we test the implementation of the radioactive
decay mechanism and the $\gamma$-transport module in our Monte Carlo radiation
hydrodynamics program. In addition, this simulation provides yet another test
for the accurate operation of the Monte Carlo radiative transfer procedures. 
Particularly, the determination of the emergent light curve is sensitive to the
correct implementation of first-order relativistic effects, such as angle
aberration and Doppler-shifts, due to the high ejecta velocities in this
application. The parameters for this test calculations, as described in
\citet{Lucy2005}, are adopted from a model SN presented in \citet{Pinto2000}. We
consider a SN with a total ejecta mass of $M = 1.39 M_{\odot}$. The ejecta are in homologous
expansion with a maximum velocity of $u_{\mathrm{max}} = 10^4 \, \mathrm{km \,
s}^{-1}$ and a uniform density of $\rho = 3.79 \times 10^{-11} \,\mathrm{g \,
cm}^{-3}$. In the ejecta, the radioactive \elem{Ni}{56} is assumed to be
strongly concentrated in the core, resulting in a \elem{Ni}{56} distribution of
$f_{\mathrm{Ni}}(M_r) = 1$ for $M_r < 0.5 \, M_{\odot}$ that linearly drops to zero
at $M_r = 0.75 \, M_{\odot}$. Here, $M_r$ denotes the mass contained within a
sphere of radius $r$. Throughout the entire material, a constant interaction
opacity for the thermal radiation field of $\chi / \rho = 0.1 \, \mathrm{cm^2 \,
g^{-1}}$ is used. Following \citet{Lucy2005}, we assume radiative equilibrium
which allows us to simplify the interaction mechanism to only include scattering
events. The grey absorption cross section for $\gamma$-radiation of
$\kappa_{\gamma} / \rho = 0.03 \, \mathrm{cm^2 \, g^{-1}}$ is adopted from
\citet{Sutherland1984} and \citet{Ambwani1988}. 

\begin{figure}
  \resizebox{\hsize}{!}{\includegraphics{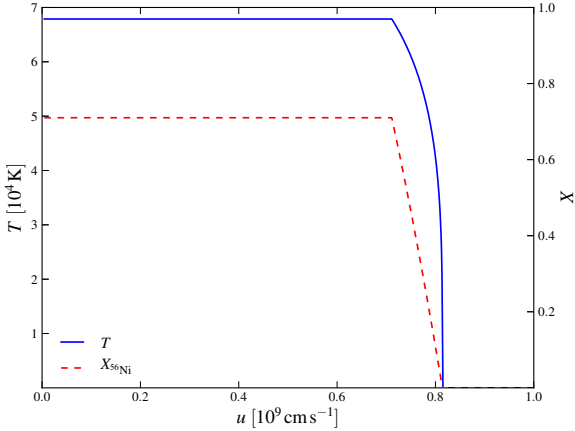}}
  \caption{Initial conditions for our Monte Carlo simulation of the
  \citet{Lucy2005} model SN ejecta at $t = 3.0 \, \mathrm{d}$. The time before
  the onset of the  simulation was bridged by an analytic homologous expansion
  calculation, resulting in the temperature profile shown as a blue solid line.
  The dashed red line describes the distribution of \elem{Ni}{56}.}
  \label{fig:lucyis}
\end{figure}

We start the Monte Carlo simulation at $t = 3.0 \, \mathrm{d}$. At much earlier
times, the high optical thickness of the ejecta material prevents an efficient
use of the Monte Carlo radiative transfer methods. We bridge the time between
the explosion to the start of the simulation by an analytic homologous expansion
calculation. Here, we assume that the entire energy released in the radioactive
decay reactions of \elem{Ni}{56} and \elem{Co}{56} immediately thermalises,
raising the gas temperature to the values shown in Fig.~\ref{fig:lucyis} at
the time of the onset of the Monte Carlo simulation.
This figure also illustrates the distribution of radioactive \elem{Ni}{56} after
the initial homologous expansion phase. Since the calculations of
\citet{Lucy2005} did not include any radiation-hydrodynamical coupling, this
interaction mechanism was also switched off in our test simulation.
Consequently, the radiative transfer is not affecting the homologous expansion
of the ejecta material.

\begin{figure}
  \resizebox{\hsize}{!}{\includegraphics{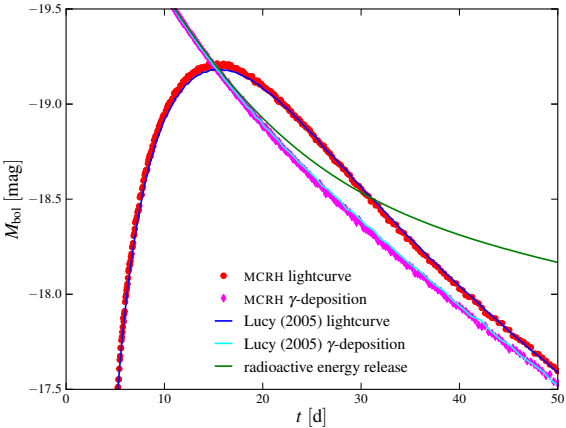}}
  \caption{Comparison between the bolometric light curve calculated with our
  Monte Carlo program (red) and the results of the \protect\citet{Lucy2005}
  calculation (blue). In addition, the rate at which the $\gamma$ radiation
  deposits its energy in the thermal radiation field is illustrated for both
  calculations (magenta and cyan). As a reference, the actual energy release in
  the decay reactions in the form of $\gamma$ radiation is also shown as the
  green line.}
  \label{fig:lucylc}
\end{figure}

Fig.~\ref{fig:lucylc} presents the bolometric light curve determined in our
Monte Carlo simulation and shows the comparison with the results of
\citet{Lucy2005}.
The excellent agreement between our simulation and the calculations performed by
\citet{Lucy2005} demonstrates the operation of the energy injection process
together with the $\gamma$-transport scheme and once more verifies the accurate
performance of our Monte Carlo approach as a whole.

\subsection{Application to the W7-Model}
\label{sec:w7}

\begin{figure}
  \resizebox{\hsize}{!}{\includegraphics{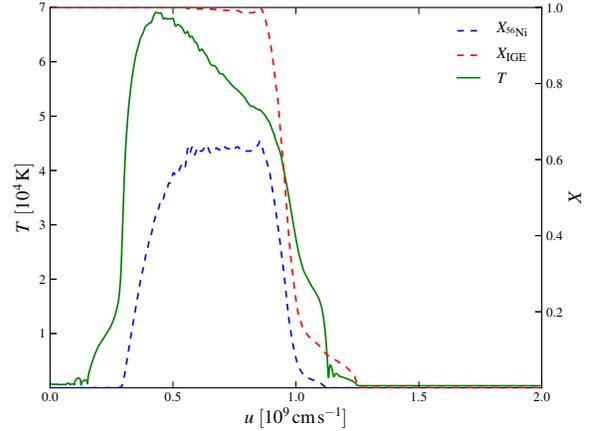}}
  \caption{The temperature profile (green line) of the W7 model after
  discretization of the original data set onto a spherical grid of 2000 cells
  and performing an homologous expansion calculation up to $t = 3 \, \mathrm{d}$
  after explosion.  In addition, the mass fraction of the radioactive
  \elem{Ni}{56} is illustrated with the dashed blue line. It illustrates the
  characteristic concentration of the \elem{Ni}{56} in the extended shell in the
  intermediate ejecta regions of the W7 model ($3 \times 10^8 \, \mathrm{cm \,
  s^{-1}} < u < 10^9 \, \mathrm{cm\, s^{-1}}$). All elements with $Z \geq 20$
  are denoted as iron group elements (IGE) whose combined distribution is shown
  as the red dashed line.}
  \label{fig:w7initial}
\end{figure}

\begin{figure}
  \resizebox{\hsize}{!}{\includegraphics{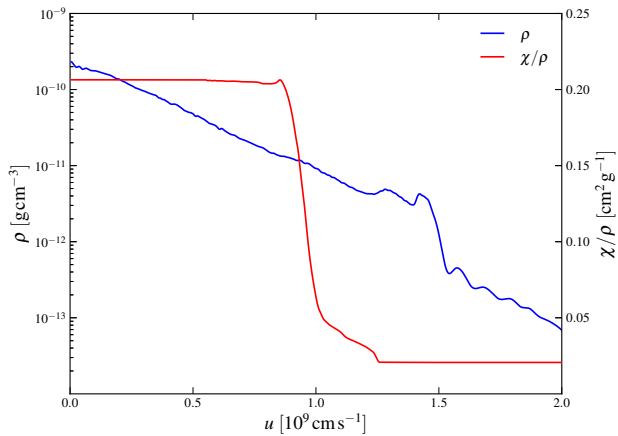}}
  \caption{The density profile of the W7 model at the start of our Monte Carlo
  simulation at $t = 3 \, \mathrm{d}$ (blue, left scale) and the scattering
  cross-section per gram (red, right scale) obtained from Equation
  (\ref{eq:sim_opacity}).}
  \label{fig:w7sigma}
\end{figure}

\begin{figure}
  \resizebox{\hsize}{!}{\includegraphics{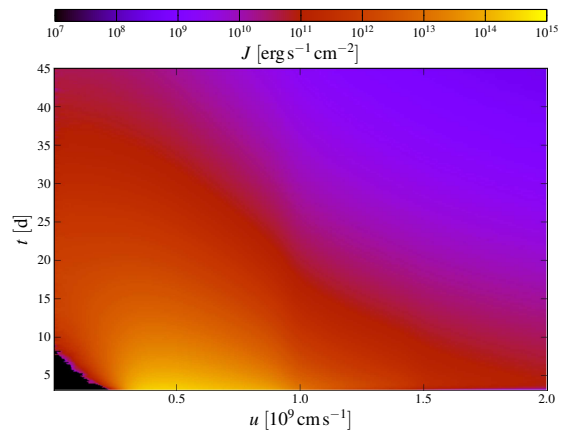}}
  \caption{Temporal evolution of the mean intensity of the radiation field in
  the radiation-hydrodynamical simulation of the W7 model. Initially, the
  radiation field is confined to the \elem{Ni}{56} shell but with decreasing
  optical depth it penetrates the inner and outer regions of the ejecta.}
  \label{fig:w7J}
\end{figure}

\begin{figure}
  \resizebox{\hsize}{!}{\includegraphics{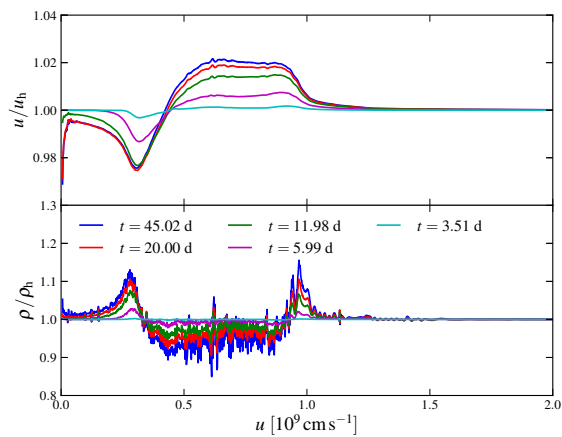}}
  \caption{Illustration of the influence of the radiation field on the evolution
  of the fluid state. In the upper panel, the velocity profile in the ejecta is
  displayed with respect to a purely homologous expansion for a variety of
  temporal snapshots (see labels in figure). In the lower panel, the density
  stratification is displayed in a corresponding fashion. The de- and
  accelerating effects of the radiation pressure are clearly visible in the
  velocity profiles, as is the resulting dilution of the \elem{Ni}{56} shell in
  the lower panel.}
  \label{fig:w7rhovel}
\end{figure}

\begin{figure}
  \resizebox{\hsize}{!}{\includegraphics{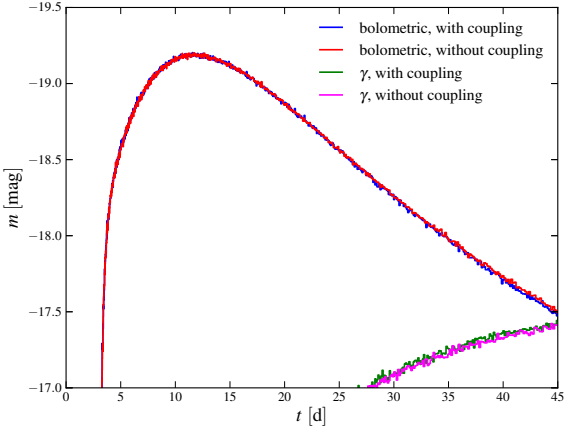}}
  \caption{Comparison between bolometric light curves calculated for the W7
  model with the radiation-hydrodynamical coupling (blue line) and with a pure
  radiative transfer calculation and homologous expansion of the ejecta material
  (red line). The emergent $\gamma$-light curve is also displayed for both
  simulations (green and cyan, respectively). No significant influence of the
  radiation-matter coupling or the resulting changes in the ejecta structure
  can be identified in the bolometric light curves.}
  \label{fig:w7lc}
\end{figure}

With the SN specific extensions, our program is well suited to examine the
influence of the radiation-hydrodynamical coupling on the observed bolometric
light curve. For this purpose we consider a SN that is described by the
well-known W7 model, first presented by \citet{Nomoto1984} and extended by larger
nuclear networks in the studies of \citet{Thielemann1986} and \citet{Iwamoto1999}.
This parameterized one-dimensional \snia{} explosion model reproduces important
observables and has thus become a standard reference in the literature.
\citet{Nomoto1984} determined the evolution of the fluid dynamical state of
the ejecta material until $t = 20 \, \mathrm{s}$ after the ignition of the
thermonuclear explosion. Characteristic for the W7 model is the concentration of
the bulk of the \elem{Ni}{56} in an extended shell, spanning the velocity region
from $3 \times 10^8 \, \mathrm{cm \, s^{-1}}$ to $10^9 \, \mathrm{cm \,
s^{-1}}$, instead of being concentrated in the core. To accurately resolve this
\elem{Ni}{56} shell, we discretize the original W7 profile to a one-dimensional
spherical grid with 2000 equidistant cells. In addition, we slightly adjust
the velocity profile of the original model to match exactly the homology
condition, allowing us to make a clear and unperturbed identification of the
influence of the radiation-matter coupling on the ejecta dynamics. As in our
calculation for the \citet{Lucy2005} toy model, we start the Monte Carlo
radiation hydrodynamics calculation at $t = 3 \, \mathrm{d}$ assuming perfect
homologous expansion behaviour of the ejecta up to this point. The resulting
temperature profile after this pure homologous phase at $t = 3 \, \mathrm{d}$ is
shown in Fig.~\ref{fig:w7initial} together with the distribution of
\elem{Ni}{56}. The density stratification at this time is visualised in
Fig.~\ref{fig:w7sigma}.
In analogy to the model simulation presented in Section \ref{sec:lucymodel} we
follow \citet{Lucy2005} and assume that the ejecta material is in radiative
equilibrium. Consequently, we can simplify the interaction mechanism between the
thermal radiation field and the surrounding material by a pure scattering
description. The grey scattering opacity is not constant throughout the ejecta
radius, but follows the distribution of heavy elements, which constitute the
dominant opacity source for the thermal radiation photons. In particular, we follow
\citet{Mazzali2006} and \citet{Sim2007} in setting the scattering opacity to
\begin{eqnarray}
  \chi / \rho = N (0.9 X_{\mathrm{IGE}} + 0.1),
  \label{eq:sim_opacity}
\end{eqnarray}
where the iron group (IGE) involves all elements heavier than calcium (the
combined distribution of these elements can be read off from
Fig.~\ref{fig:w7initial}).  The scaling factor $N$ is chosen to ensure a mean
interaction cross section of $\langle \chi/\rho \rangle = 0.1 \, \mathrm{cm^2 \,
g^{-1}}$. Fig.~\ref{fig:w7sigma} shows the effect of the heavy elements on the
interaction cross section throughout the ejecta material.
All interactions of the $\gamma$-radiation are described by a single constant
absorption opacity of $\kappa_{\gamma} / \rho = 0.03 \, \mathrm{cm^2 \, g^{-1}}$
(see Section~\ref{sec:lucymodel}). With these parameters, the temporal evolution
of the hydrodynamical state of the W7 ejecta is calculated up to $t = 45 \,
\mathrm{d}$. Fig.~\ref{fig:w7J} illustrates the behaviour of the radiation
field, quantified by the mean intensity $J$, during this period of time.
At the beginning, the radiation field is concentrated at the location of the
\elem{Ni}{56}-shell, where initially most of the $\gamma$-interactions take
place due to the high optical depth of the ejecta. As time progresses, the
ejecta become increasingly transparent to the radiation field, which begins to
penetrate the inner regions and propagates to the outer edge of the supernova ejecta.
Here, the Monte Carlo packets escape and are recorded to determine the
bolometric light curve.  Due to the radiation pressure, the initially confined
radiation field leaves imprints on the velocity and density profiles of the
ejecta material as it propagates in- and outwards.  The changes in the ejecta
structure are indicated in Fig.~\ref{fig:w7rhovel},
which shows the velocity and density profiles with respect to a purely homologous
expansion that would occur in the absence of the radiation-hydrodynamical
coupling. The radiation pressure accelerates the ejecta motion in the outer
parts of the \elem{Ni}{56} shell and decelerates the expansion in the inner
regions.  Both effects dilute the \elem{Ni}{56} shell and pile up material at
the edges of this region. The influence of the radiation pressure is 
strongest in the early phases of the expansion, where the high optical depth
causes a strong coupling of the radiation field to the surrounding material and
stalls as the ejecta become more and more transparent. In total, the radiation
pressure induces deviations from the purely homologous density profile on the
order of 10 per cent, which are compatible with the findings of the previous study by
\citet{Woosley2007}. The structural changes in the density stratification are,
however, not prominent enough to significantly alter the shape of the emergent
bolometric light curve. Neglecting the radiation-hydrodynamical coupling and
assuming a purely homologous expansion yields nearly the same bolometric light curve as a
detailed calculation. A comparison of both simulations is shown in
Fig.~\ref{fig:w7lc}.
Even if the bolometric light curve remains unaffected by changes in the fluid
state, colour light curves may be affected since the radiation pressure changes
the velocity of ejecta regions in which different elements are concentrated.
However, this effect cannot be studied with the current grey implementation of
our approach to radiation hydrodynamics and remains to be re-addressed with a
chromatic version of the code.

All simulations presented in this section required about $3.5 \times 10^{13}$
floating point operations.  In these calculations the initial radiation field at
$t = 3 \, \mathrm{d}$ was discretized by $10^6$ Monte Carlo packets. Due to the
energy release in the decay reactions and the outflow of radiation packets at
the outer edge of the ejecta, the number of active packets varied greatly during
the simulation, but a maximum of $2.43 \times 10^6$ packets were used to represent
the radiation field at any time. 

\begin{figure}
  \resizebox{\hsize}{!}{\includegraphics{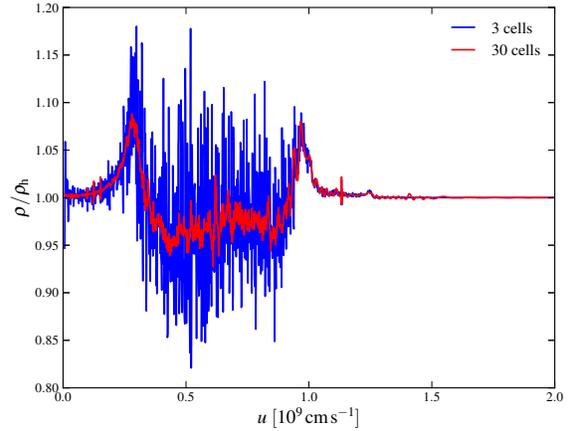}}
  \caption{Illustration of the influence of smoothing on the density evolution
  of the ejecta. The density profiles at $t = 14.3 \, \mathrm{d}$ are displayed
  with respect to a purely homologous expansion (cf.\ Fig~.\ref{fig:w7rhovel}).
  We show the results of two simulations: one that employs smoothing over the
  neighbouring 30 cells (red line) and one with very mild smoothing over 3
  cells (blue line). As this comparison shows, the smoothing approach removes the
  strong stochastic cell-to-cell fluctuations, but retains the overall physical
  variation.}
  \label{fig:w7smooth}
\end{figure}

Despite the large number of packets, the Monte Carlo estimators for the
radiation force are subject to a significant level of statistical noise. In this
particular case, the entire energy-momentum transfer is determined by the
radiative flux in the co-moving frame (cf.\ Equations
\ref{eq:radf0_analytic_cmf} and \ref{eq:radf3_analytic_cmf}). As pointed out in
Sections \ref{sec:discretization} and \ref{sec:noise}, our discretization
approach is not ideally suited to accurately capture this quantity in regions
where the radiation field is nearly isotropic, such as the inner parts of the
\elem{Ni}{56} shell. However, despite being subject to a considerable level of
Monte Carlo noise, the radiative flux captures the expected effect of the
radiation field inflating the \elem{Ni}{56} bubble (see Figure
\ref{fig:w7smooth}).  Thus, we employed a smoothing kernel (cf.\ Section
\ref{sec:noise}) in our simulations, averaging over 30 neighbouring cells and
thereby avoiding difficulties in the hydrodynamical solver. In the absence of
this smoothing, cell-to-cell fluctuations in the fluid state lead to convergence
problems in the Riemann solver that could not be averted by increasing the
number of active packets on computationally feasible scales.  Using the
smoothing capability gives a significant reduction of the cell-to-cell
fluctuations without damaging the physical variations in the quantities, as
Fig.~\ref{fig:w7smooth} illustrates.  Here, the density stratification at $t =
14.3 \, \mathrm{d}$ is shown for both a simulation that includes smoothing over
30 neighbouring cells and for one with very mild smoothing over 3 cells only,
proving that the approach preserves the physical signal.

\section{Discussion}
\label{sec:discussion}

In this paper we presented a Monte Carlo approach to radiation hydrodynamical
problems in astrophysical environments. By combining Monte Carlo radiative
transfer methods that rely on the indivisible packet formalism
\citep{Abbott1985} with the finite-volume hydrodynamical technique PPM
\citep{Colella1984} we have aimed to retain the benefits of the Monte Carlo
machinery for the modelling of complex interaction physics and arbitrary
geometries. Here, our main focus lay on the development and presentation of the
necessary numerical tools and on demonstrating the operation of this method, its
physical accuracy and its computational feasibility. By using volume-based Monte
Carlo estimators \citep{Lucy1999, Lucy2002, Lucy2003} in the reconstruction of
radiation field characteristics, the maximum amount of information is extracted
from the propagation behaviour of Monte Carlo packets and the Monte Carlo noise
is minimized. 

A series of toy calculations has been performed to test the operation of the
main components of our Monte Carlo radiation hydrodynamical method. In
particular, the simulation of radiative shocks verified the accuracy of our
approach to a standard radiation hydrodynamical test.  As expected, due to
the nature of the Monte Carlo method, calculations in optically thick
environments are time-consuming but feasible and accurate, as the radiative
shock examples showed. In general, all calculations were completed within hours
to a day on a single desktop CPU.  However, due to the very efficient scaling
behaviour of the Monte Carlo algorithm to large numbers of processor cores, a
future parallel implementation of the method provides the scope for 
significant decreases in the run time. 

The application to \snia{} ejecta successfully demonstrated the operation of our
method to an astrophysical problem for which this method was primarily
developed. In this exercise, the influence of the radiation-matter coupling on
the density stratification during the near-homologous expansion phase of the ejecta
has been investigated. The results we obtained are in agreement with the
previous study by \citet{Woosley2007} who used the radiation hydrodynamics code
\textsc{stella} \citep{Blinnikov2006}. The induced changes in the ejecta
structure, however, were confirmed to have no significant influence on the bolometric
light curve, as predicted by \citet{Pinto2000}.

Despite the agreement of our results with previous studies, the \snia{}
application also illustrated some difficulties of our approach. Our
discretization scheme into packets of radiative energy allows us to easily
construct all relevant radiation field characteristics and can be generalized to
a fully frequency-dependent transport treatment in a straight-forward manner.
However, in regions where the radiation field is close to LTE or to isotropy,
the radiation force components are subject to considerable statistical
fluctuations due to this discretization choice. Here, we suppressed this noise
component by applying a smoothing kernel. Although beyond the scope of this
work, in the future further reduction of the statistical noise should be
explored by incorporating implicit Monte Carlo techniques
\citep{Fleck1971} or a Monte Carlo radiative transfer approach that
is based on the difference formulation \citep{Szoke2005, Brooks2005}. For such
schemes, it will be important to consider how all the physical processes
necessary to adequately address particular astrophysical applications can be implemented.

As the main aim of this work was to establish the methods and the numerical
framework of the Monte Carlo radiation hydrodynamics approach, all calculations
were performed with a simplified implementation of the method. In particular, we
restricted our tests to one-dimensional geometries and the radiative transfer
was performed in a grey approximation with a very simple opacity prescription.
However, we stress that these simplifications were made only to reduce the
complexity and the computational effort of the test simulations. They do not
affect the applicability and operation of the Monte Carlo radiation
hydrodynamical approach itself. In the future, we aim to introduce more
sophisticated opacity prescriptions and make the transition from grey transport
to a fully frequency-dependent radiative transfer scheme. This will add a
further level of sophistication to the method, but will not impact the operation
of the radiation-matter coupling in our approach. The tools to realise the
frequency-dependent transfer have already been developed and are provided for
example in the framework of the macro-atom formalism by \citet{Lucy2002}. In
addition to improving the physical accuracy of the radiative transfer, our
long-term efforts will be directed towards a generalised implementation for
multi-dimensional problems. With this generalization our radiation
hydrodynamical method will include all major capabilities that have already made
the Monte Carlo technique a very successful and rewarding approach for pure
radiation transport applications.

\section*{Acknowledgements}
  UMN wishes to thank K.~S.~Long, for introducing him to the mysteries and
  beauties of Monte Carlo radiative transfer techniques.  Furthermore, UMN is
  very grateful to E. M\"uller for fruitful discussions on interfacing
  deterministic and probabilistic approaches and to P.~Edelmann for providing
  the PPM implementation. The authors also thank the anonymous reviewer for his
  valuable comments, which helped to improve the paper.  UMN, SAS, FKR and MK
  acknowledge financial support by the Group of Eight/Deutscher Akademischer
  Austauch Dienst (Go8/DAAD) Australian-German Joint Research Co-operation
  Scheme (``Supernova explosions: comparing theory with observations.''). The
  work of FKR was supported by the Deutsche Forschungsgemeinschaft via the Emmy
  Noether Program (RO 3676/1-1) and by the ARCHES prize of the German Federal
  Ministry of Education and Research (BMBF).

\end{document}